\documentclass[
    11pt,
    letterpaper,
    reprint,
    notitlepage,
    superscriptaddress,
    aps,prx
]{revtex4-2}

\usepackage{amsmath,amssymb,amsfonts} % standard AMS packages
\usepackage{mathrsfs}
\usepackage{empheq}
\usepackage{physics}

\usepackage[normalem]{ulem}

\usepackage{microtype}

\usepackage[svgnames,dvipsnames]{xcolor}
\usepackage{graphicx}
\definecolor{NewBlue}{rgb}{0.1, 0.1, 0.7}
\definecolor{NewRed}{rgb}{0.7, 0.1, 0.1}
\usepackage[colorlinks,
    linkcolor=Maroon,
    citecolor=NewBlue,
    urlcolor=ForestGreen]{hyperref}

\usepackage[capitalise]{cleveref}

\usepackage[
    exponent-product=\cdot,
    range-phrase=\text{--},
    range-units=single,
    separate-uncertainty,
]{siunitx}[=v2]

\usepackage{tikz-feynman}
\usepackage{pdfpages}
\makeatletter
\AtBeginDocument{\let\LS@rot\@undefined}
\makeatother

\renewcommand{\t}[1]{\mathrm{{#1}}}

\renewcommand{\phi}{\varphi}

\newcommand{\avg}[1]{\left \langle {#1}\right \rangle}

\newcommand{\oprho}{\hat \rho}

\newcommand{\p}{\partial}

\newcommand{\LigoMIT}{LIGO Laboratory, Massachusetts Institute of Technology, Cambridge, MA 02139}
\newcommand{\MechMIT}{Department of Mechanical Engineering, Massachusetts Institute of Technology, Cambridge, MA 02139}

\newcommand{\PhysHarv}{Department of Physics, Harvard University, Cambridge, MA 02138}

\begin{document}

\title{Distinguishable consequence of classical gravity on quantum matter}
\author{Serhii Kryhin}
\affiliation{\PhysHarv}
\author{Vivishek Sudhir}
\affiliation{\LigoMIT}
\affiliation{\MechMIT}
\date{\today}

\begin{abstract}
What if gravity is classical? If true, a consistent co-existence of classical gravity and quantum matter 
requires that gravity exhibit irreducible fluctuations. These fluctuations can mediate
classical correlations, but not quantum entanglement, between the quantized motion of the gravitationally interacting matter.
We use a consistent theory of quantum-classical dynamics in the Newtonian limit of gravity to show
that experimentally relevant observables can conclusively test the hypothesis that gravity is
classical. This can be done for example by letting highly coherent source masses interact with each other gravitationally, and performing precise measurements of the cross-correlation of their motion.
Theory predicts a characteristic phase response that distinguishes classical gravity from quantum gravity,
and from naive sources of decoherence. Such experiments are imminently viable.
\end{abstract}

\maketitle

\emph{Introduction.} 
It is believed that if gravitational source masses can be prepared in quantum superpositions, 
then the gravitational field sourced by them has to be quantum. 
Feynman's argument in support \cite{Feyn57} relies on the expectation that
a double-slit experiment with massive particles produces an interference pattern.
If the gravitational fields sourced by them are assumed to be classical, 
then it can convey which-path information that contradicts the development
of the interference pattern. So for consistency, the gravitational field needs to be quantum, so that its quantum
fluctuations obfuscate the which-path information (see also \cite{BelAsp18,DanWald22}).
An equally viable hypothesis that achieves consistency is that the gravitational field is
classical and stochastic \cite{HuVerd08,TilDio16,Opp18,OppZach22,Oppenheim:2022xjr}, 
so that it cannot convey precise which-path information. 

Contrary to the hypothesis that gravity is quantum \cite{Carn22,BoseMaz22}, 
stochastic classical gravity should not entangle the source masses. 
But since the gravitational field is a dynamical entity, it can mediate classical correlations
between the motion of the source masses. 
It is precisely this subtle detail that is missed 
by naively adopting the view that the only effect of classical stochastic gravity is to 
decohere the source masses \cite{karolyhazy_gravitation_1966,
Dio87,penrose_gravitys_1996,Anast96,blencowe_effective_2013,Kafri14,Bass17,kanno_noise_2021,
anastopoulos_gravitational_2022}.
It certainly does, but it leaves a telltale sign distinct from quantum gravity, and from other extraneous
sources of decoherence.

We describe a consistent and generic low-energy theory of classical gravity interacting with quantum 
source masses, and produce experimentally relevant signatures of such a theory. 
Consideration of a concrete experimental scenario allows us to explicitly verify the lack of entanglement
in such theories. Crucially however, the correlations mediated by the gravitational field
% is smoking-gun evidence for the hypothesis that gravity is classical.
% These are potential smoking gun signature for the hypothesis that 
% gravity is classical. 
are sensitive to the difference between a classical description of gravity 
and an effective quantum theory of gravitation 
\cite{Brons12,Gupta52,Feyn63,Schw63,deWitt67,Wein64ii,Don94}, and can be tested without macroscopic
masses in quantum superpositions \cite{Bose17,MarVed17}.

\emph{Classical-quantum theory. }
Even if gravity is classical, it is necessary that when a quantized mass interacts with it, 
the ensuing dynamics does not prevent the assignment of a legitimate joint state.  
That is, there exists a positive-definite unit-trace operator $\hat{\rho}_t(z)$ in the Hilbert space
of the quantum object for each classical state $z$ in the phase space of the classical gravitational field. 
The (quantum) state of the mass alone is $\hat{\rho}_Q = \int \dd{z} \hat{\rho}_t (z)$, while
the (classical) state of the field alone is $p_t(z) = \Tr \hat{\rho}_t (z)$.
The general structure of the dynamics of $\hat{\rho}_t(z)$ is \cite{Aleks81,BouTras88,PrezKis97,PerTer01,
Opp18,OppZach22,Oppenheim:2022xjr}
\begin{equation}\label{eq:rhoGeneral}
	\begin{split}
	\dot{\hat{\rho}}_t = 
	&-i[\hat{H}_t,\hat{\rho}_t] 
		+ Q_{\alpha \beta} \left(\hat{L}_\alpha \hat{\rho}_t \hat{L}_\beta^\dagger 
	  	-\frac{1}{2} [\hat{L}_\beta^\dagger \hat{L}_\alpha, \hat{\rho}_t]_+ \right) \\
	&+\p_{z_i} \left( C_{i} \hat \rho_t \right)  +  \p_{z_i z_j} \left( C_{ij} \hat \rho_t \right) \\
	&+\p_{z_i} \left( M_{\alpha i} \hat\rho_t \hat L^\dagger_\alpha + \t{h.c.} \right).
\end{split}
\end{equation}
This equation consists of several qualitatively distinct terms: 
the first row describes pure quantum state evolution by the Hamiltonian $\hat{H}_t(z)$
and quantum diffusion described by the Lindblad operators $\{\hat{L}_\alpha\}$ with diffusion 
constants $Q_{\alpha \beta}(z)$; the second row, as we shall see, describes classical pure state evolution and
diffusion with constants $C_i(z)$ and $C_{ij}(z)$; and the last line is ``classical-quantum diffusion'' 
with constants $M_{\alpha i}(z)$. 

Note that the classical-quantum model of \cref{eq:rhoGeneral} describes the co-evolution of
quantum matter and classical gravity via their mutual back-action. Thus, it
resolves the pathologies of a semi-classical description of gravity \cite{AnHu14}, and prior experiments
where quantum systems evolve in a static gravitational potential \cite{PouReb60,cow75,PetChu01,Nesv02,Jenk11,
RosTino14,AseKas17} have not tested this model.

The state of the classical system $p_t(z)$ obeys classical Hamiltonian dynamics under the Hamiltonian
$H_C$ if $\p_{z_i} C_i = 0$ and $C_j + J_{ij}\p_{z_i} H_C =0$; here $J$ is the symplectic matrix. 
Under these conditions, $\Tr [\p_{z_i}(C_i \hat \rho)] = J_{ij} (\p_{z_i}H_C)(\p_{z_j}p) = \{H_C, p\}$,
where $\{\cdot,\cdot\}$ is the Poisson bracket.
If a similar set of conditions are imposed on the coefficients $M_{\alpha i}$, namely that,
$\p_{z_i} M_{\alpha i}=0$ and $M_{\alpha j} + J_{ij} (\p_{z_i} h_\alpha) =0$, for some function $h_\alpha$, then
$\t{Tr} [\p_{z_i}(M_{\alpha i}\hat{\rho}\hat L_\alpha^\dagger + \t{h.c.})] = \Tr \{\hat{H}_I,\hat{\rho}\}$, where
$\hat{H}_I = h_\alpha \hat{L}_\alpha^\dagger + \t{h.c.}$ is a Hermitian operator. By considering the case where the
matter is classical, and eliminating it to obtain equations of motion for the state of the gravitational field alone
(see Appendix I), it can be shown that $\hat{H}_I$ is an interaction Hamiltonian.
So far, only the trace-ful part of the classical-quantum diffusion has been accounted for. Explicitly separating out its
trace-free part (see Appendix~II) gives rise to an additional function $H'$ which, 
together with $H_I$, subsumes the classical-quantum diffusion term.
In sum, \cref{eq:rhoGeneral} reduces to:
\begin{equation}\label{eq:rhoWithHamiltonian}
	\begin{split}
	\dot{\hat{\rho}}_t &= -i[\hat{H},\hat{\rho}_t] +\, Q_{\alpha \beta} \left(\hat{L}_\alpha \hat{\rho}_t\hat{L}_\beta^\dagger 
	  -\frac{1}{2} [\hat{L}_\beta^\dagger \hat{L}_\alpha, \hat{\rho}_t]_+ \right) +
	\\
	 &\t{Herm} \left\{ H, \hat \rho \right\} 
	+ i \, \t{AntiHerm} \{ \hat H^\prime, \hat \rho_t \}
	+ \p_{z_i z_j} \left( C_{ij} \hat \rho_t(z) \right),
	\end{split}
\end{equation}
where $\hat H = \hat H_Q + H_C + \hat H_I$, $\t{Herm}/\t{AntiHerm}$ denote the hermitian and anti-hermitian parts, and $H'$ is a
new function, independent of the Hamiltonian $H$, that will turn out to describe the irreducible effect of 
the classical system on the quantum system.

We will now describe, within the formalism above,
a consistent theory of classical gravity interacting with quantum masses, such that the gravitational 
interaction reduces to the (experimentally relevant) Newtonian limit.

\emph{The Hamiltonian $\hat{H}$.} 
The Newtonian gravitational potential $\Phi$ is sourced by the mass density $f$ via
$\nabla^2 \Phi = -4\pi G f$. This derives from the Lagrangian $L = \int  
[-\abs{\nabla \Phi}^2/(8\pi G) + \Phi f] \dd^3 x$.
Since in Newtonian gravity, the potential is instantaneously determined by the source, the conjugate momentum 
$\Pi = \delta L/\delta \dot{\Phi} = 0$. This (primary) constraint makes the Lagrangian singular and
passage to the Hamiltonian subtle. Dirac's theory of constrained dynamical
systems \cite{Dirac_1950,Dirac58} can nevertheless be deployed to construct a modified Hamiltonian
(see Appendix III.A for the details). The result is 
$H = \int [\abs{\nabla \Phi}^2/(8\pi G) - \Phi f + \lambda \Pi] \dd^3 x$, where $\lambda$ is the Lagrange
multiplier that is completely determined by enforcing the primary constraint. The Hamilton equation, $\dot{\Pi} = \nabla^2 \Phi/(4\pi G)
+f = 0$, reproduces the Newtonian field equation. The term $\lambda \Pi$ simply enforces the constraint
via the other Hamilton equation $\dot{\Phi}=\lambda$; it is thus
dynamically irrelevant.
By promoting the mass density to a quantum operator, we arrive at
\begin{equation} \label{eq:Hz}
	H_C(z) + \hat H_I = \int \dd^3 x \left[\lambda \Pi + \frac{|\nabla \Phi|^2}{8 \pi G} - \Phi \hat f(x) \right].
\end{equation}
The first two terms are purely classical and therefore correspond to $H_C$, and the last term has both quantum and classical degrees of freedom, and therefore corresponds to $\hat H_I$.
(A Hamiltonian of a similar form had been divined previously \cite{Di_si_2011}. The above Hamiltonian can also be
derived by carefully treating the Newtonian limit of
general relativity, see Appendix III.B for details of development from standard literature \cite{carroll2003spacetime}. 
In that approach, which is different from the canonical formalism of general relativity \cite{Dir59,ADM59}, 
the Lagrangian is non-degenerate, and so the $\lambda \Pi$ term does not appear.)

Importantly, and in contrast to semi-classical theories of quantum gravity, the above construction deals with
the mass density operator $\hat{f}$, and not its expectation value --- thus eluding the pathologies of the
semi-classical theory. 
The full Hamiltonian is $\hat{H}(z) = \hat{H}_Q + {H}_C(z) + \hat H_I $, where $\hat{H}_Q$ is the
Hamiltonian of the quantized matter.

\emph{The operator $\hat{H}'$ and quantum-classical diffusion.} Now we turn attention to the new function 
$\hat H^\prime (z)$ that appears in \cref{eq:rhoWithHamiltonian}. 
Even though $\hat H^\prime$ produces evolution of $\hat \rho_t(z)$, it turns out to have no influence on the dynamics of the 
classical field alone. 
Thus the form of $\hat H^\prime$ is intrinsic to the classical-quantum theory and has to be supplied
by this theory. As we now show, its structure can be fixed from knowledge of the relevant degrees of freedom at play.

For the gravitational field, the relevant degree of freedom in the Newtonian limit 
is the potential $\Phi$ (consistent with linearized gravity, see Appendix III and \footnote{Other gauge choices, such as in Ref. \cite{LayOpp23}, give rise to other degrees of freedom; however
in the weak-field adiabatic limit, agreement between the different approaches is manifest.}). 
For the quantized matter it is its mass density $\hat{f}(x)$.
Therefore, by dimensional analysis, to the lowest order in $1/c$, the function $H'$ can only be
\begin{equation}
	\hat H^\prime (z) = \epsilon \int \dd^3 x \; \Phi(x) \hat f (x),
\end{equation}
where $\epsilon$ is a dimensionless constant that needs to be experimentally determined. 
% Note that the form of $\hat H^\prime$ is similar to the interaction term in the Hamiltonian in \cref{eq:Hz}. 
% Note that $\hat H^\prime$ has to have both classical and quantum components, because purely classical 
% or purely quantum term will nulify the combination of the Poisson bracket and antihermitization.

The remaining terms in \cref{eq:rhoWithHamiltonian}, proportional to $Q_{\alpha \beta}$ and 
$C_{ij}$, turn out to be irrelevant as far as weak local gravity is concerned. 
As described above, in the weak-gravity regime, the relevant phase-space variables of the field are 
$z=(\Phi,\Pi=0)$; to preserve locality, we assume that $C_{ij}(z) = c_{ij}(z) \delta(x_i -x_j)$ and
$Q_{\alpha \beta} =q_{\alpha \beta}(z) \delta(x_\alpha - x_\beta)$. Under these assumptions, explicit calculation shows
(see Appendix IV) that the lowest non-trivial contribution ---
from the second order term in $z$ --- simply renormalizes the phenomenological constant $\epsilon$.
The zeroth order term, clearly independent of gravity, can be neglected. Thus, the relevant contribution is
the first-order effect conveyed by $H'$.

\emph{Dynamics of quantum matter.}
Suppose we assume that it is only the quantized matter that is accessible for experiments. Then the above
theory can only be tested vis-\'{a}-vis the properties encoded in the state
% The goal of the computation is to produce a theory of classical gravity of quantum matter that can be used to obtain concrete experimaental predictions. With this direction in hand, we claim that the full equation of the form of \cref{eq:rhoWithHamiltonian} is no need to us, since we \emph{cannot} directly observe $\Phi(x)$ in a real experiment. A realistic experiment is inherently always destined to be only able to probe for the effects that classicality of $\Phi$ imposes on the quantum degrees of freedom. Therefore, we find it useful to imtegrate out the classical degrees of freedom and work with a fully quantum density matrix
$\hat \rho_Q (t) \equiv \int \dd{z} \hat \rho_t(z).$
% \begin{equation}
% 	\hat \rho_Q (t) \equiv \int \dd{z} \hat \rho_t(z).
% \end{equation}
In order to derive an equation of motion for this quantum state, it is necessary to eliminate
the gravitational potential $\Phi$ from \cref{eq:rhoWithHamiltonian}.
This is facilitated by the observation that the gravity Hamiltonian [\cref{eq:Hz}] together with 
the classical-quantum model implies the modified Newtonian law (see Appendix IV):
\begin{equation}\label{eq:QuantumNewtonsLaw}
\begin{split}
	\int \dd{z} \nabla_x^2 \Phi(x) \hat{\rho}_t(z) 
	= -4\pi G & \left(  \frac{1}{2}[\hat{f}(x),\hat{\rho}_Q]_+ \right. \\
		&\left. \quad + i\epsilon [\hat{f}(x),\hat{\rho}_Q] \right),
\end{split}
\end{equation}
where $[\cdot,\cdot]_+$ is the anti-commutator. (Note that term involving $\lambda$ ends up being irrelevant for the field dynamics, as expected. Exactly the same result follows from the approach rooted in general relativity, in Appendix IV.B, 
as a consequence of the Newtonian limit of the theory, and not an additional assumption \cite{Di_si_2011}.) 
Substituting \cref{eq:QuantumNewtonsLaw} in \cref{eq:rhoGeneral} and integrating out the classical gravitational
degree of freedom gives (restoring $\hbar$)
\begin{equation} \label{eq:LindbladFinal}
\begin{split}
	\dot{\hat{\rho}}_Q = 
	&-\frac{i}{\hbar} [\hat{H}_\t{eff}, \hat{\rho}_Q] \\
	&- \epsilon \frac{G}{2\hbar} \int \frac{\dd^3 x \, \dd^3 y}{|x - y|} \left[ \hat f(x),  \left[ \hat f(y), 
		\hat \rho_Q \right] \right] \\
	&+ Q^{\alpha \beta} \left( \hat L_\alpha \hat \rho_Q \hat L_\beta^\dagger 
		- \frac{1}{2} \left[ \hat L_\alpha^\dagger \hat L_\beta, \hat \rho_Q \right]_+ \right),
\end{split}
\end{equation}
which is a closed Lindblad equation for $\hat \rho_Q$. 
$\hat{H}_\t{eff} = \hat{H}_Q + \hat{H}_G$ is the effective Hamiltonian, where
\begin{equation}\label{eq:HG}
	\hat H_G = -\frac{G}{2} \int \dd^3 x \, \dd^3 x^\prime \; 
	\frac{\hat f (x) \hat f(x^\prime)}{|x - x^\prime|},
\end{equation}
arises from the anti-commutator term in \cref{eq:QuantumNewtonsLaw}, and is $z$-independent. 
For a collection of point masses, $\hat{f}(x) = \sum_i m_i \delta(x-\hat{x}_i)$, and so $\hat{H}_G$
describes the quantized Newtonian interaction between them. 
The stochastic effect of classical gravity is contained in the Lindblad term proportional to $\epsilon$
in \cref{eq:LindbladFinal} --- its origin is the anti-commutator term in \cref{eq:QuantumNewtonsLaw}, which
traces back to the function $H'(z)$. 
In contrast to past work \cite{Di_si_2011,OppZach22,LayOpp23}, we \emph{derive} the
gravity-dependent Lindbladian term from the natural structure of the dynamics of 
the joint state, and the consistency of that with gravity in the Newtonian regime. This Lindbladian
arises from the function $H'$, and not from a weak-field expansion of $Q^{\alpha \beta}$ or $C_{ij}$
\footnote{This means that the constraints derived in Refs. \cite{OppZach22,LayOpp23}, based 
on positive-definiteness of the evolution, are inapplicable to $\epsilon$.}.
In our approach, since $Q^{\alpha \beta}$ is assumed local, to lowest order, 
it is $\epsilon$-independent, and so
the corresponding Lindbladian terms simply describe additional decoherence of the quantum matter.

In sum, $\epsilon = 0$ reduces to a quantum description of gravity in the Newtonian
limit \cite{TilDio16,Anastopoulos_2021,blencowe_effective_2013}, where matter evolves unitarily under Newtonian potential 
(decoherence from quantum fluctuations in gravity are sub-leading in $G$, see Appendix V).
Complementarily, $\epsilon \neq 0$
corresponds to classical gravity, where an additional fluctuation in the gravitational field of the same order of magnitude as Newtonian potential affects the quantum matter. 
Notably, these fluctuations are correlated across the matter degrees of freedom. 

\emph{Form of the gravity Lindbladian revisited.}
Consider the example of two localized objects of masses $m_{1,2}$ 
separated by a distance $R$. Suppose the quantum fluctuations in the displacement of the two masses
$\hat{x}_{1,2}$ are small compared to $R$, then the gravity Lindbladian in \cref{eq:LindbladFinal}
\begin{equation*}
	\mathcal{L}_G[\hat{\rho}_Q] \equiv - \epsilon \frac{G}{2\hbar} \int \frac{\dd^3 x \, \dd^3 x'}{|x - x'|} 
		\left[ \hat f(x), \left[ \hat f(x'), \hat \rho_Q \right] \right]
\end{equation*}
reduces to the simple form
\begin{equation}\label{eq:LGfromQC}
	\mathcal{L}_G[\hat{\rho}_Q] \approx - \epsilon \frac{G m_1 m_2}{\hbar R^3} 
	[ \hat x_1, [ \hat x_2, \hat \rho_Q  ] ].
\end{equation}
This
form of the Lindbladian is in fact completely natural in the Newtonian limit in the sense that
it can be derived from purely dimensional arguments once the relevant degrees of freedom are known. To wit,
assuming that the classical description of gravity only depends on the position of source masses, and
that its effect is translation invariant, the Lindbladian can be taken to be
of the form 
\begin{equation}\label{eq:LG}
	\mathcal{L}_G[\oprho_Q] = \frac{Gm_1 m_2}{\hbar R}[\hat{L}_1(\hat{\ell}_1 -\hat{\ell_2}), 
		[\hat{L}_2(\hat{\ell}_1 -\hat{\ell}_2),\oprho_Q]], 
\end{equation} 
where $\hat{L}_i$ are dimensionless Hermitian operators that depend on 
the difference of $\hat{\ell}_i = \hat{x}_i/R$. 
The dimensional pre-factor is fixed since it is the only suitable combination relevant in the low energy
limit of concern here. (Other combinations, such as $\hbar/(G m^3 R) ,\sqrt{G/(c^3 R^2)}$, come up only in higher 
orders in $R$, which are irrelevant in the Newtonian limit.) 
Since the quantum fluctuations of the source masses are small, we only consider,
$\hat{L}_i(\hat{\ell}_1 - \hat{\ell}_2) = \epsilon_i (\hat{\ell}_1 - \hat{\ell}_2)$, where
the proportionality constant is arbitrary (but real). 
Inserting this in \cref{eq:LG} gives three terms: two of the form (no sum on $i$) 
$\epsilon_i^2 [\hat{\ell}_i [\hat{\ell}_i,\oprho_Q]]$; 
and another of the form $\epsilon_1 \epsilon_2 [\hat{\ell}_1,[\hat{\ell}_2,\oprho_Q]]$.
The former describes gravitational decoherence of either mass independent of the other; 
it may thus be absorbed into the description of the thermal bath that any realistic source mass 
will invariably be coupled to.
The latter term describes joint decoherence of the masses due to their mutual back-action mediated
via the stochastic classical field. This term,
$\mathcal{L}_G[\hat \rho_Q] \approx (\epsilon_1 \epsilon_2) (G m_1 m_2/R^3)
[\hat{x}_1,[\hat{x}_2, \hat \rho_Q]]$, is 
precisely the form derived in \cref{eq:LGfromQC}. 

Note that the decoherence described by $\mathcal{L}_G$ is distinct from decoherence of gravitating
bodies in a fully quantum description of linearized gravity that arise from vacuum 
fluctuations of the gravitational field. 
In particular, the effect of decoherence due to graviton emission by the two bodies will be local
and at order $G^2$, i.e. described by a Lindbladian of the form (no sum on $i$) 
$\sim G^2 [\hat{x}_i,[\hat{x}_i,\hat{\rho}_Q]]$.
Both the $G^2$ scaling and the form can be understood in analogy with radiation reaction in quantum 
electrodynamics: decoherence due to spontaneous emission is second-order in the fine-structure constant, and
each emitter --- as long as they are separated by more than the wavelength of the emitted radiation ---
decoheres independent of the other. Clearly, the joint decoherence described by $\mathcal{L}_G$ is
qualitatively and quantitatively distinct from the expectation of linearized quantum gravity.

\begin{figure}[t!]
\centering
\includegraphics[width=\columnwidth]{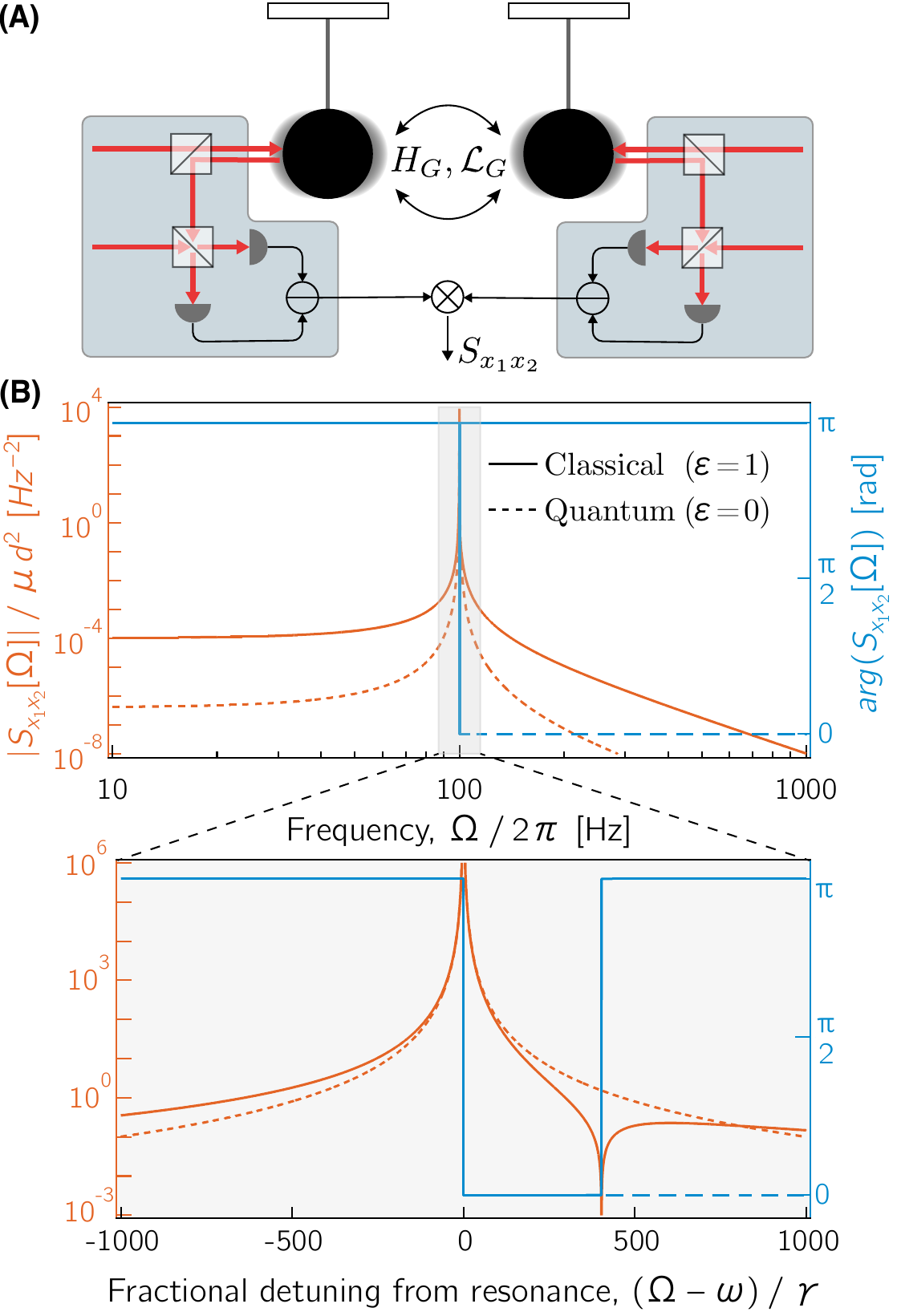}
\caption{\label{fig:Sxx}
	(A) Cartoon of an experimental arrangement to measure the effect of classical gravity on quantum masses.
	A pair of masses (black) are suspended as harmonic oscillators, which interact with each other 
	gravitationally, described by the Hamiltonian $\hat{H}_G$ [\cref{eq:HG}] and the 
	Lindbladian $\mathcal{L}_G$ [\cref{eq:LGfromQC}]. 
	A pair of interferometers (gray) measure their displacements 
	continuously; the records of these measurements are used to estimate the cross-correlation between 
	their displacement fluctuations $S_{x_1 x_2}$.
	(B) The theoretical prediction for the cross-correlation $S_{x_1 x_2}$.
	Here the label ``Quantum'' corresponds to setting $\epsilon = 0$, in which case, gravity is a quantum
	field; ``Classical" corresponds to $\epsilon =1$. 
	The two oscillators are assumed identical with resonance frequency $\omega = 2\pi\cdot 100\, \t{Hz}$, and
	damping rate $\gamma = 2\pi\cdot 10^{-3}\, \t{Hz}$.
	The orange traces depict $\abs{S_{x_1 x_2}}$ while blue depicts the phase $\arg S_{x_1 x_2}$.
	The effect of the gravitational Lindbladian is an increase in the correlation far from resonance (by a factor 
	$\omega / 2 \gamma (2 \bar N +1)$ compared to the quantum case), and an additional $\pi$ phase shift
	at a detuning $\Delta \Omega \approx 2 \gamma (2 \bar N +1)$ from resonance (see zoomed plot below).
}
\end{figure}

\emph{Experimental signature.} We now consider an experimentally relevant example where two massive
quantum harmonic oscillators interact with each other gravitationally (see \Cref{fig:Sxx}(A)). 
As we will show, if gravity is assumed classical (i.e. $\epsilon > 0$), their motion is correlated via the 
classical stochastic gravitational field mediating their interaction, and 
these correlations offer a qualitatively distinct prediction from the assumption that gravity is 
quantum (i.e. $\epsilon =0$).

We take the masses to be one-dimensional quantum harmonic oscillators of frequency $\omega_{1,2}$
each coupled to independent thermal baths of occupation $\bar{N}_{1,2}$ at rates $\gamma_{1,2}$.
They are also gravitationally coupled to each other by their close proximity of distance $R$. 
This scenario is described by the equation
\begin{equation}\label{eq:master_equation}
	\dot{\rho}_Q(t) = -\frac{i}{\hbar} [\hat H_\t{eff},\hat \rho_Q ] 
	+ \mathcal{L}\left[\hat \rho_Q \right],
\end{equation}
where $\hat{H}_\t{eff} = \hat{H}_Q + \hat{H}_G$, 
with $\hat{H}_Q = \sum_i \hbar \omega_i (\hat{a}_i^\dagger \hat{a}_i +\tfrac{1}{2})$, and 
$\mathcal{L} = \mathcal{L}_G + \sum_i \mathcal{L}_i$, where $\mathcal{L}_i [\hat \rho_Q] 
= \gamma_i(\bar N_i + 1)
(\hat a_i \hat \rho_Q \hat a^\dagger_i - \tfrac{1}{2} [\hat{a}_i^\dagger \hat{a}_i,\oprho_Q]_+ )
+\gamma_i \bar N_i ( \hat a^\dagger_i \hat \rho_Q \hat a_i 
- \tfrac{1}{2} [\hat a_i \hat a^\dagger_i,\oprho_Q]_+ )$, describes the realistic coupling of any experimental
source mass to a thermal bath of average thermal occupation $\bar{N}_i$ \cite{gardiner00}.
Here the gravitational interaction Hamiltonian $\hat{H}_G \approx G m_1 m_2 \hat{x}_1\hat{x}_2/(2 R^3)$ is
obtained by inserting the mass density $\hat{f}(x) = m_1 \delta(x+\tfrac{R}{2}-\hat{x}_1) 
+ m_2 \delta(x-\tfrac{R}{2}-\hat{x}_2)$ in \cref{eq:HG}, and then performing a small-distance expansion of 
the Coulombic denominator and the resulting integrals. 
(Note that here we have dropped a singular self-interaction term whose origin is the point-mass
approximation of the center-of-mass degree of freedom of the oscillator; for an oscillator that is formed of an elastic continuum, 
these terms will excite internal modes.)

The primary effect of the gravitational interaction is the correlated decoherence of the two masses
via the gravity Lindbladian $\mathcal{L}_G$. Simply put, the motion of one mass causes motion in the other mass
via their gravitational interaction; however, since classical gravity is necessarily noisy, the induced motion
reflects this fact. When the two masses are in equilibrium --- with their respective thermal environment and
with each other --- 
the simplest experimentally accessible observable sensitive to non-zero value of $\epsilon$ 
is the the cross-correlation $\avg{\hat{x}_1(t) \hat{x}_2(0)}$ of the quantized 
displacements $\hat{x}_{1,2}$ of the two gravitating oscillators. 
This can be inferred from the outcomes of continuous measurements of their displacements, say by
interferometry. (Such measurements cause back-action \cite{BragKhal92}
whose effect is an increased apparent temperature \cite{PurReg13a,Teuf16}, which can be absorbed 
into the bath occupations $\bar{N}_i$; 
back-action-evading measurements may also be imagined \cite{SuSchw14,LecTeu15,YapCorb19}.)

The theory presented above can produce a concrete prediction for the 
cross-correlation spectrum of the displacement  $S_{x_1 x_2}[\Omega] = \frac{1}{2} \int_{-\infty}^{+\infty} dt \, e^{- i \Omega t}\langle [\hat x_1(t); \hat x_2(0)]_+\rangle$. We do so by first
mapping the master equation [\cref{eq:master_equation}] into a partial differential equation (PDE) for 
the characteristic function $\chi(\vb{\xi},\vb{\xi}^*,t) = \t{Tr} [\oprho_Q(t) \hat{D}(\vb{\xi})]$,
where $\hat{D}(\xi) = \prod_i \t{exp}[\xi_i \hat{a}_i^\dagger - \xi_i^* \hat{a}_i]$ are a set of 
complete orthogonal basis in the set of operators of the joint Hilbert space 
of the two oscillators \cite{CahGlau69o}.
Since the resulting PDE is of second order, and the initial thermal state of the oscillators is Gaussian 
in $\chi(\xi,\xi^*,0)$, a Gaussian ansatz suffices to solve for $\chi(\xi,\xi^*,t)$.
From this solution, we obtain equal-time correlations of the coordinates and momenta of the oscillators. 
Finally, employing the quantum regression theorem \cite{gardiner00}, we get the unequal-time 
correlations, whose Fourier transform gives the cross-correlation spectrum 
(see Appendix VI for the gory details). 
For identical oscillators, the result is (Appendix VI contains the full expressions)
\begin{equation}
	S_{x_1 x_2}[\Omega] \approx -\frac{x_\t{zp}^2}{\gamma_G}
	\begin{dcases}
		\epsilon + \frac{2\bar{N}+1}{Q}; & \Omega \ll \omega \\
		\left( \frac{\omega}{\Omega} \right)^4 \left( \epsilon - \frac{2\bar{N}+1}{Q} \right); & \Omega \gg \omega, 	
	\end{dcases} 
\end{equation}
where, $Q = \omega/(2\gamma)$ is the quality factor of the oscillator, $x_\t{zp}$ its zero-point motion,
and $\gamma_G = 2Gm/(R^3\omega)$.
Importantly, gravitationally interacting oscillators with quality factor 
$Q \gtrsim \bar{N}/\epsilon$ can be sensitive probes
of whether $\epsilon$ is zero or not. 

\Cref{fig:Sxx}(B) depicts the cross-correlation spectrum $S_{x_1 x_2}$ for a typical scenario of two 
identical mechanical oscillators coupled via gravity, under the hypothesis that gravity is 
quantum ($\epsilon=0$) or that it is a classical stochastic field, with $\epsilon =1$ chosen arbitrarily.
In the quantum case, the correlation between the two oscillators is purely via the quantized 
Newtonian interaction Hamiltonian $\hat{H}_G$, resulting in correlated motion across all frequencies,
with most of the motion concentrated around the common resonance. 
In the classical stochastic case, the joint interaction mediated by $\mathcal{L}_G$ apparently produces
anti-correlated motion away from resonance, accompanied by a characteristic phase shift
$\arg S_{x_1 x_2}$. This phase shift persists for \emph{all} non-zero values of 
$\epsilon$ (unlike the anti-correlated feature in $\abs{S_{x_1 x_2}}$, which can diminish with 
$\epsilon$) and is thus a robust experimental signature that gravity is classical.
Thus, careful measurements of the cross-correlation of two gravitationally coupled highly coherent
mechanical oscillators can distinguish between the hypothesis that gravity is classical or quantum.

The concrete scenario of two massive oscillators, also allows us to explicitly verify that the 
quantum-classical theory of gravity does not mediate entanglement 
(see Appendix VII where we employ the Gaussian PPT criterion \cite{Simon00,Duan00} to test for entanglement) 
--- a natural expectation \cite{Kafri14} that has not been verified so far in a consistent dynamical theory. Indeed, the criterion that classical gravitational fluctuations produce no entanglement implies that $\epsilon \geq 8\gamma/(3\gamma_G)$.

\emph{Conclusion. }We have derived a natural and pathology-free theory of two quantized bodies interacting 
via classical gravity, and produced an experimentally testable implication of this theory that 
can distinguish whether gravity is classical or not. Testing this prediction calls for an experiment 
where two highly coherent massive oscillators are coupled to each 
other via gravity. Importantly, they do not need to be in superposition 
states \cite{Bose17,MarVed17}. Further, the experimental signature in the cross-correlation of the
motion of the two masses is a feature distinct from the 
effects of garden-variety decoherence: masses in close proximity to each other
are susceptible to a host of nefarious near-field forces which cause decoherence described by
\emph{local} Lindbladians of the form (no sum over $i$) 
$[\hat{x}_i,[\hat{x}_i,\hat{\rho}_Q]]/R^n$, for $n\geq 4$ --- importantly they do not cause joint 
(or correlated) decoherence as described by $\mathcal{L}_G$. 
Such joint decoherence can 
arise from a common third entity the masses may interact with; isolating from such interactions is the 
art of the experimentalist. 
But such careful experiments can falsify the premise that gravity is 
classical by hunting for the signatures we have derived.
In this sense, the current proposal differs from bounding extraneous decoherence 
of a single massive body \cite{Oppenheim:2022xjr} or violation of an ``LOCC inequality'' \cite{Lami24}.

\emph{Acknowledgement.} We thank Antoine Tilloy for a valuable discussion. V.S. was partially supported by the Class of 1957 Career Development Chair. 

\bibliography{refs_CQ_gravity}

\begin{thebibliography}{61}%
\makeatletter
\providecommand \@ifxundefined [1]{%
 \@ifx{#1\undefined}
}%
\providecommand \@ifnum [1]{%
 \ifnum #1\expandafter \@firstoftwo
 \else \expandafter \@secondoftwo
 \fi
}%
\providecommand \@ifx [1]{%
 \ifx #1\expandafter \@firstoftwo
 \else \expandafter \@secondoftwo
 \fi
}%
\providecommand \natexlab [1]{#1}%
\providecommand \enquote  [1]{``#1''}%
\providecommand \bibnamefont  [1]{#1}%
\providecommand \bibfnamefont [1]{#1}%
\providecommand \citenamefont [1]{#1}%
\providecommand \href@noop [0]{\@secondoftwo}%
\providecommand \href [0]{\begingroup \@sanitize@url \@href}%
\providecommand \@href[1]{\@@startlink{#1}\@@href}%
\providecommand \@@href[1]{\endgroup#1\@@endlink}%
\providecommand \@sanitize@url [0]{\catcode `\\12\catcode `\$12\catcode
  `\&12\catcode `\#12\catcode `\^12\catcode `\_12\catcode `\%12\relax}%
\providecommand \@@startlink[1]{}%
\providecommand \@@endlink[0]{}%
\providecommand \url  [0]{\begingroup\@sanitize@url \@url }%
\providecommand \@url [1]{\endgroup\@href {#1}{\urlprefix }}%
\providecommand \urlprefix  [0]{URL }%
\providecommand \Eprint [0]{\href }%
\providecommand \doibase [0]{https://doi.org/}%
\providecommand \selectlanguage [0]{\@gobble}%
\providecommand \bibinfo  [0]{\@secondoftwo}%
\providecommand \bibfield  [0]{\@secondoftwo}%
\providecommand \translation [1]{[#1]}%
\providecommand \BibitemOpen [0]{}%
\providecommand \bibitemStop [0]{}%
\providecommand \bibitemNoStop [0]{.\EOS\space}%
\providecommand \EOS [0]{\spacefactor3000\relax}%
\providecommand \BibitemShut  [1]{\csname bibitem#1\endcsname}%
\let\auto@bib@innerbib\@empty
%</preamble>
\bibitem [{\citenamefont {Feynman}(1957)}]{Feyn57}%
  \BibitemOpen
  \bibfield  {author} {\bibinfo {author} {\bibfnamefont {R.~P.}\ \bibnamefont
  {Feynman}},\ }\bibfield  {title} {\bibinfo {title} {The necessity of
  gravitational quantization},\ }in\ \href@noop {} {\emph {\bibinfo {booktitle}
  {The role of {Gravitation} in {Physics}: {Report} from the 1957 {Chapel}
  {Hill} conference}}}\ (\bibinfo {year} {1957})\BibitemShut {NoStop}%
\bibitem [{\citenamefont {Belenchia}\ \emph {et~al.}(2018)\citenamefont
  {Belenchia}, \citenamefont {Wald}, \citenamefont {Giacomini}, \citenamefont
  {Castro-Ruiz}, \citenamefont {Brukner},\ and\ \citenamefont
  {Aspelmeyer}}]{BelAsp18}%
  \BibitemOpen
  \bibfield  {author} {\bibinfo {author} {\bibfnamefont {A.}~\bibnamefont
  {Belenchia}}, \bibinfo {author} {\bibfnamefont {R.~M.}\ \bibnamefont {Wald}},
  \bibinfo {author} {\bibfnamefont {F.}~\bibnamefont {Giacomini}}, \bibinfo
  {author} {\bibfnamefont {E.}~\bibnamefont {Castro-Ruiz}}, \bibinfo {author}
  {\bibfnamefont {{\v C}.}~\bibnamefont {Brukner}},\ and\ \bibinfo {author}
  {\bibfnamefont {M.}~\bibnamefont {Aspelmeyer}},\ }\bibfield  {title}
  {\bibinfo {title} {Quantum superposition of massive objects and the
  quantization of gravity},\ }\href
  {https://doi.org/10.1103/PhysRevD.98.126009} {\bibfield  {journal} {\bibinfo
  {journal} {Physical Review D}\ }\textbf {\bibinfo {volume} {98}},\ \bibinfo
  {pages} {126009} (\bibinfo {year} {2018})}\BibitemShut {NoStop}%
\bibitem [{\citenamefont {Danielson}\ \emph {et~al.}(2022)\citenamefont
  {Danielson}, \citenamefont {Satishchandran},\ and\ \citenamefont
  {Wald}}]{DanWald22}%
  \BibitemOpen
  \bibfield  {author} {\bibinfo {author} {\bibfnamefont {D.~L.}\ \bibnamefont
  {Danielson}}, \bibinfo {author} {\bibfnamefont {G.}~\bibnamefont
  {Satishchandran}},\ and\ \bibinfo {author} {\bibfnamefont {R.~M.}\
  \bibnamefont {Wald}},\ }\bibfield  {title} {\bibinfo {title} {Gravitationally
  mediated entanglement: {Newtonian} field versus gravitons},\ }\href
  {https://doi.org/10.1103/PhysRevD.105.086001} {\bibfield  {journal} {\bibinfo
   {journal} {Physical Review D}\ }\textbf {\bibinfo {volume} {105}},\ \bibinfo
  {pages} {086001} (\bibinfo {year} {2022})}\BibitemShut {NoStop}%
\bibitem [{\citenamefont {Hu}\ and\ \citenamefont
  {Verdaguer}(2008)}]{HuVerd08}%
  \BibitemOpen
  \bibfield  {author} {\bibinfo {author} {\bibfnamefont {B.~L.}\ \bibnamefont
  {Hu}}\ and\ \bibinfo {author} {\bibfnamefont {E.}~\bibnamefont {Verdaguer}},\
  }\bibfield  {title} {\bibinfo {title} {Stochastic {Gravity}: {Theory} and
  {Applications}},\ }\href {https://doi.org/10.12942/lrr-2008-3} {\bibfield
  {journal} {\bibinfo  {journal} {Living Reviews in Relativity}\ }\textbf
  {\bibinfo {volume} {11}},\ \bibinfo {pages} {3} (\bibinfo {year}
  {2008})}\BibitemShut {NoStop}%
\bibitem [{\citenamefont {Tilloy}\ and\ \citenamefont
  {Di{\'o}si}(2016)}]{TilDio16}%
  \BibitemOpen
  \bibfield  {author} {\bibinfo {author} {\bibfnamefont {A.}~\bibnamefont
  {Tilloy}}\ and\ \bibinfo {author} {\bibfnamefont {L.}~\bibnamefont
  {Di{\'o}si}},\ }\bibfield  {title} {\bibinfo {title} {Sourcing semiclassical
  gravity from spontaneously localized quantum matter},\ }\href
  {https://doi.org/10.1103/PhysRevD.93.024026} {\bibfield  {journal} {\bibinfo
  {journal} {Physical Review D}\ }\textbf {\bibinfo {volume} {93}},\ \bibinfo
  {pages} {024026} (\bibinfo {year} {2016})}\BibitemShut {NoStop}%
\bibitem [{\citenamefont {Oppenheim}(2023)}]{Opp18}%
  \BibitemOpen
  \bibfield  {author} {\bibinfo {author} {\bibfnamefont {J.}~\bibnamefont
  {Oppenheim}},\ }\bibfield  {title} {\bibinfo {title} {A postquantum theory of
  classical gravity?},\ }\href {https://doi.org/10.1103/PhysRevX.13.041040}
  {\bibfield  {journal} {\bibinfo  {journal} {Phys. Rev. X}\ }\textbf {\bibinfo
  {volume} {13}},\ \bibinfo {pages} {041040} (\bibinfo {year}
  {2023})}\BibitemShut {NoStop}%
\bibitem [{\citenamefont {Oppenheim}\ and\ \citenamefont
  {Weller-Davies}(2022)}]{OppZach22}%
  \BibitemOpen
  \bibfield  {author} {\bibinfo {author} {\bibfnamefont {J.}~\bibnamefont
  {Oppenheim}}\ and\ \bibinfo {author} {\bibfnamefont {Z.}~\bibnamefont
  {Weller-Davies}},\ }\bibfield  {title} {\bibinfo {title} {The constraints of
  post-quantum classical gravity},\ }\href
  {https://doi.org/10.1007/JHEP02(2022)080} {\bibfield  {journal} {\bibinfo
  {journal} {Journal of High Energy Physics}\ }\textbf {\bibinfo {volume}
  {2022}},\ \bibinfo {pages} {80} (\bibinfo {year} {2022})}\BibitemShut
  {NoStop}%
\bibitem [{\citenamefont {Oppenheim}\ \emph {et~al.}(2023)\citenamefont
  {Oppenheim}, \citenamefont {Sparaciari}, \citenamefont {\v{S}oda},\ and\
  \citenamefont {Weller-Davies}}]{Oppenheim:2022xjr}%
  \BibitemOpen
  \bibfield  {author} {\bibinfo {author} {\bibfnamefont {J.}~\bibnamefont
  {Oppenheim}}, \bibinfo {author} {\bibfnamefont {C.}~\bibnamefont
  {Sparaciari}}, \bibinfo {author} {\bibfnamefont {B.}~\bibnamefont
  {\v{S}oda}},\ and\ \bibinfo {author} {\bibfnamefont {Z.}~\bibnamefont
  {Weller-Davies}},\ }\bibfield  {title} {\bibinfo {title} {{Gravitationally
  induced decoherence vs space-time diffusion: testing the quantum nature of
  gravity}},\ }\href {https://doi.org/10.1038/s41467-023-43348-2} {\bibfield
  {journal} {\bibinfo  {journal} {Nature Commun.}\ }\textbf {\bibinfo {volume}
  {14}},\ \bibinfo {pages} {7910} (\bibinfo {year} {2023})}\BibitemShut
  {NoStop}%
\bibitem [{\citenamefont {Carney}(2022)}]{Carn22}%
  \BibitemOpen
  \bibfield  {author} {\bibinfo {author} {\bibfnamefont {D.}~\bibnamefont
  {Carney}},\ }\bibfield  {title} {\bibinfo {title} {Newton, entanglement, and
  the graviton},\ }\href {https://doi.org/10.1103/PhysRevD.105.024029}
  {\bibfield  {journal} {\bibinfo  {journal} {Physical Review D}\ }\textbf
  {\bibinfo {volume} {105}},\ \bibinfo {pages} {024029} (\bibinfo {year}
  {2022})}\BibitemShut {NoStop}%
\bibitem [{\citenamefont {Bose}\ \emph {et~al.}(2022)\citenamefont {Bose},
  \citenamefont {Mazumdar}, \citenamefont {Schut},\ and\ \citenamefont {Toro{\v
  s}}}]{BoseMaz22}%
  \BibitemOpen
  \bibfield  {author} {\bibinfo {author} {\bibfnamefont {S.}~\bibnamefont
  {Bose}}, \bibinfo {author} {\bibfnamefont {A.}~\bibnamefont {Mazumdar}},
  \bibinfo {author} {\bibfnamefont {M.}~\bibnamefont {Schut}},\ and\ \bibinfo
  {author} {\bibfnamefont {M.}~\bibnamefont {Toro{\v s}}},\ }\bibfield  {title}
  {\bibinfo {title} {Mechanism for the quantum natured gravitons to entangle
  masses},\ }\href {https://doi.org/10.1103/PhysRevD.105.106028} {\bibfield
  {journal} {\bibinfo  {journal} {Physical Review D}\ }\textbf {\bibinfo
  {volume} {105}},\ \bibinfo {pages} {106028} (\bibinfo {year}
  {2022})}\BibitemShut {NoStop}%
\bibitem [{\citenamefont {Karolyhazy}(1966)}]{karolyhazy_gravitation_1966}%
  \BibitemOpen
  \bibfield  {author} {\bibinfo {author} {\bibfnamefont {F.}~\bibnamefont
  {Karolyhazy}},\ }\bibfield  {title} {\bibinfo {title} {Gravitation and
  quantum mechanics of macroscopic objects},\ }\href
  {https://doi.org/10.1007/BF02717926} {\bibfield  {journal} {\bibinfo
  {journal} {Il Nuovo Cimento A}\ }\textbf {\bibinfo {volume} {42}},\ \bibinfo
  {pages} {390} (\bibinfo {year} {1966})}\BibitemShut {NoStop}%
\bibitem [{\citenamefont {Di{\'o}si}(1987)}]{Dio87}%
  \BibitemOpen
  \bibfield  {author} {\bibinfo {author} {\bibfnamefont {L.}~\bibnamefont
  {Di{\'o}si}},\ }\bibfield  {title} {\bibinfo {title} {A universal master
  equation for the gravitational violation of quantum mechanics},\ }\href
  {https://doi.org/10.1016/0375-9601(87)90681-5} {\bibfield  {journal}
  {\bibinfo  {journal} {Physics Letters A}\ }\textbf {\bibinfo {volume}
  {120}},\ \bibinfo {pages} {377} (\bibinfo {year} {1987})}\BibitemShut
  {NoStop}%
\bibitem [{\citenamefont {Penrose}(1996)}]{penrose_gravitys_1996}%
  \BibitemOpen
  \bibfield  {author} {\bibinfo {author} {\bibfnamefont {R.}~\bibnamefont
  {Penrose}},\ }\bibfield  {title} {\bibinfo {title} {On {Gravity}'s role in
  {Quantum} {State} {Reduction}},\ }\href {https://doi.org/10.1007/BF02105068}
  {\bibfield  {journal} {\bibinfo  {journal} {General Relativity and
  Gravitation}\ }\textbf {\bibinfo {volume} {28}},\ \bibinfo {pages} {581}
  (\bibinfo {year} {1996})}\BibitemShut {NoStop}%
\bibitem [{\citenamefont {Anastopoulos}(1996)}]{Anast96}%
  \BibitemOpen
  \bibfield  {author} {\bibinfo {author} {\bibfnamefont {C.}~\bibnamefont
  {Anastopoulos}},\ }\bibfield  {title} {\bibinfo {title} {Quantum theory of
  nonrelativistic particles interacting with gravity},\ }\href
  {https://doi.org/10.1103/PhysRevD.54.1600} {\bibfield  {journal} {\bibinfo
  {journal} {Physical Review D}\ }\textbf {\bibinfo {volume} {54}},\ \bibinfo
  {pages} {1600} (\bibinfo {year} {1996})}\BibitemShut {NoStop}%
\bibitem [{\citenamefont {Blencowe}(2013)}]{blencowe_effective_2013}%
  \BibitemOpen
  \bibfield  {author} {\bibinfo {author} {\bibfnamefont {M.~P.}\ \bibnamefont
  {Blencowe}},\ }\bibfield  {title} {\bibinfo {title} {Effective {Field}
  {Theory} {Approach} to {Gravitationally} {Induced} {Decoherence}},\ }\href
  {https://doi.org/10.1103/PhysRevLett.111.021302} {\bibfield  {journal}
  {\bibinfo  {journal} {Physical Review Letters}\ }\textbf {\bibinfo {volume}
  {111}},\ \bibinfo {pages} {021302} (\bibinfo {year} {2013})}\BibitemShut
  {NoStop}%
\bibitem [{\citenamefont {Kafri}\ \emph {et~al.}(2014)\citenamefont {Kafri},
  \citenamefont {Taylor},\ and\ \citenamefont {Milburn}}]{Kafri14}%
  \BibitemOpen
  \bibfield  {author} {\bibinfo {author} {\bibfnamefont {D.}~\bibnamefont
  {Kafri}}, \bibinfo {author} {\bibfnamefont {J.~M.}\ \bibnamefont {Taylor}},\
  and\ \bibinfo {author} {\bibfnamefont {G.~J.}\ \bibnamefont {Milburn}},\
  }\bibfield  {title} {\bibinfo {title} {A classical channel model for
  gravitational decoherence},\ }\href
  {https://doi.org/10.1088/1367-2630/16/6/065020} {\bibfield  {journal}
  {\bibinfo  {journal} {New Journal of Physics}\ }\textbf {\bibinfo {volume}
  {16}},\ \bibinfo {pages} {065020} (\bibinfo {year} {2014})}\BibitemShut
  {NoStop}%
\bibitem [{\citenamefont {Bassi}\ \emph {et~al.}(2017)\citenamefont {Bassi},
  \citenamefont {Gro{\ss}ardt},\ and\ \citenamefont {Ulbricht}}]{Bass17}%
  \BibitemOpen
  \bibfield  {author} {\bibinfo {author} {\bibfnamefont {A.}~\bibnamefont
  {Bassi}}, \bibinfo {author} {\bibfnamefont {A.}~\bibnamefont
  {Gro{\ss}ardt}},\ and\ \bibinfo {author} {\bibfnamefont {H.}~\bibnamefont
  {Ulbricht}},\ }\bibfield  {title} {\bibinfo {title} {Gravitational
  decoherence},\ }\href {http://stacks.iop.org/0264-9381/34/i=19/a=193002}
  {\bibfield  {journal} {\bibinfo  {journal} {Classical and Quantum Gravity}\
  }\textbf {\bibinfo {volume} {34}},\ \bibinfo {pages} {193002} (\bibinfo
  {year} {2017})}\BibitemShut {NoStop}%
\bibitem [{\citenamefont {Kanno}\ \emph {et~al.}(2021)\citenamefont {Kanno},
  \citenamefont {Soda},\ and\ \citenamefont {Tokuda}}]{kanno_noise_2021}%
  \BibitemOpen
  \bibfield  {author} {\bibinfo {author} {\bibfnamefont {S.}~\bibnamefont
  {Kanno}}, \bibinfo {author} {\bibfnamefont {J.}~\bibnamefont {Soda}},\ and\
  \bibinfo {author} {\bibfnamefont {J.}~\bibnamefont {Tokuda}},\ }\bibfield
  {title} {\bibinfo {title} {Noise and decoherence induced by gravitons},\
  }\href {https://doi.org/10.1103/PhysRevD.103.044017} {\bibfield  {journal}
  {\bibinfo  {journal} {Physical Review D}\ }\textbf {\bibinfo {volume}
  {103}},\ \bibinfo {pages} {044017} (\bibinfo {year} {2021})}\BibitemShut
  {NoStop}%
\bibitem [{\citenamefont {Anastopoulos}\ and\ \citenamefont
  {Hu}(2022)}]{anastopoulos_gravitational_2022}%
  \BibitemOpen
  \bibfield  {author} {\bibinfo {author} {\bibfnamefont {C.}~\bibnamefont
  {Anastopoulos}}\ and\ \bibinfo {author} {\bibfnamefont {B.-L.}\ \bibnamefont
  {Hu}},\ }\bibfield  {title} {\bibinfo {title} {Gravitational decoherence: {A}
  thematic overview},\ }\href {https://doi.org/10.1116/5.0077536} {\bibfield
  {journal} {\bibinfo  {journal} {AVS Quantum Science}\ }\textbf {\bibinfo
  {volume} {4}},\ \bibinfo {pages} {015602} (\bibinfo {year}
  {2022})}\BibitemShut {NoStop}%
\bibitem [{\citenamefont {Bronstein}(2012)}]{Brons12}%
  \BibitemOpen
  \bibfield  {author} {\bibinfo {author} {\bibfnamefont {M.}~\bibnamefont
  {Bronstein}},\ }\bibfield  {title} {\bibinfo {title} {Republication of:
  {Quantum} theory of weak gravitational fields},\ }\href
  {https://doi.org/10.1007/s10714-011-1285-4} {\bibfield  {journal} {\bibinfo
  {journal} {General Relativity and Gravitation}\ }\textbf {\bibinfo {volume}
  {44}},\ \bibinfo {pages} {267} (\bibinfo {year} {2012})}\BibitemShut
  {NoStop}%
\bibitem [{\citenamefont {Gupta}(1952)}]{Gupta52}%
  \BibitemOpen
  \bibfield  {author} {\bibinfo {author} {\bibfnamefont {S.~N.}\ \bibnamefont
  {Gupta}},\ }\bibfield  {title} {\bibinfo {title} {{Quantization of Einstein's
  Gravitational Field: General Treatment}},\ }\href
  {https://doi.org/10.1088/0370-1298/65/8/304} {\bibfield  {journal} {\bibinfo
  {journal} {Proceedings of the Physical Society A}\ }\textbf {\bibinfo
  {volume} {65}},\ \bibinfo {pages} {608} (\bibinfo {year} {1952})}\BibitemShut
  {NoStop}%
\bibitem [{\citenamefont {Feynman}(1963)}]{Feyn63}%
  \BibitemOpen
  \bibfield  {author} {\bibinfo {author} {\bibfnamefont {R.~P.}\ \bibnamefont
  {Feynman}},\ }\bibfield  {title} {\bibinfo {title} {{Quantum theory of
  gravitation}},\ }\href@noop {} {\bibfield  {journal} {\bibinfo  {journal}
  {Acta Phys. Polon.}\ }\textbf {\bibinfo {volume} {24}},\ \bibinfo {pages}
  {697} (\bibinfo {year} {1963})}\BibitemShut {NoStop}%
\bibitem [{\citenamefont {Schwinger}(1963)}]{Schw63}%
  \BibitemOpen
  \bibfield  {author} {\bibinfo {author} {\bibfnamefont {J.}~\bibnamefont
  {Schwinger}},\ }\bibfield  {title} {\bibinfo {title} {Quantized
  {Gravitational} {Field}},\ }\href {https://doi.org/10.1103/PhysRev.130.1253}
  {\bibfield  {journal} {\bibinfo  {journal} {Physical Review}\ }\textbf
  {\bibinfo {volume} {130}},\ \bibinfo {pages} {1253} (\bibinfo {year}
  {1963})}\BibitemShut {NoStop}%
\bibitem [{\citenamefont {DeWitt}(1967)}]{deWitt67}%
  \BibitemOpen
  \bibfield  {author} {\bibinfo {author} {\bibfnamefont {B.~S.}\ \bibnamefont
  {DeWitt}},\ }\bibfield  {title} {\bibinfo {title} {{Quantum Theory of
  Gravity. I. The Canonical Theory}},\ }\href
  {https://doi.org/10.1103/PhysRev.160.1113} {\bibfield  {journal} {\bibinfo
  {journal} {Physical Review}\ }\textbf {\bibinfo {volume} {160}},\ \bibinfo
  {pages} {1113} (\bibinfo {year} {1967})}\BibitemShut {NoStop}%
\bibitem [{\citenamefont {Weinberg}(1964)}]{Wein64ii}%
  \BibitemOpen
  \bibfield  {author} {\bibinfo {author} {\bibfnamefont {S.}~\bibnamefont
  {Weinberg}},\ }\bibfield  {title} {\bibinfo {title} {Feynman {Rules} for
  {Any} {Spin}. {II}. {Massless} {Particles}},\ }\href
  {https://doi.org/10.1103/PhysRev.134.B882} {\bibfield  {journal} {\bibinfo
  {journal} {Physical Review}\ }\textbf {\bibinfo {volume} {134}},\ \bibinfo
  {pages} {B882} (\bibinfo {year} {1964})}\BibitemShut {NoStop}%
\bibitem [{\citenamefont {Donoghue}(1994)}]{Don94}%
  \BibitemOpen
  \bibfield  {author} {\bibinfo {author} {\bibfnamefont {J.~F.}\ \bibnamefont
  {Donoghue}},\ }\bibfield  {title} {\bibinfo {title} {General relativity as an
  effective field theory: The leading quantum corrections},\ }\href
  {https://doi.org/10.1103/PhysRevD.50.3874} {\bibfield  {journal} {\bibinfo
  {journal} {Phys. Rev. D}\ }\textbf {\bibinfo {volume} {50}},\ \bibinfo
  {pages} {3874} (\bibinfo {year} {1994})}\BibitemShut {NoStop}%
\bibitem [{\citenamefont {Bose}\ \emph {et~al.}(2017)\citenamefont {Bose},
  \citenamefont {Mazumdar}, \citenamefont {Morley}, \citenamefont {Ulbricht},
  \citenamefont {Toro{\v s}}, \citenamefont {Paternostro}, \citenamefont
  {Geraci}, \citenamefont {Barker}, \citenamefont {Kim},\ and\ \citenamefont
  {Milburn}}]{Bose17}%
  \BibitemOpen
  \bibfield  {author} {\bibinfo {author} {\bibfnamefont {S.}~\bibnamefont
  {Bose}}, \bibinfo {author} {\bibfnamefont {A.}~\bibnamefont {Mazumdar}},
  \bibinfo {author} {\bibfnamefont {G.~W.}\ \bibnamefont {Morley}}, \bibinfo
  {author} {\bibfnamefont {H.}~\bibnamefont {Ulbricht}}, \bibinfo {author}
  {\bibfnamefont {M.}~\bibnamefont {Toro{\v s}}}, \bibinfo {author}
  {\bibfnamefont {M.}~\bibnamefont {Paternostro}}, \bibinfo {author}
  {\bibfnamefont {A.~A.}\ \bibnamefont {Geraci}}, \bibinfo {author}
  {\bibfnamefont {P.~F.}\ \bibnamefont {Barker}}, \bibinfo {author}
  {\bibfnamefont {M.}~\bibnamefont {Kim}},\ and\ \bibinfo {author}
  {\bibfnamefont {G.}~\bibnamefont {Milburn}},\ }\bibfield  {title} {\bibinfo
  {title} {Spin {Entanglement} {Witness} for {Quantum} {Gravity}},\ }\href
  {https://doi.org/10.1103/PhysRevLett.119.240401} {\bibfield  {journal}
  {\bibinfo  {journal} {Physical Review Letters}\ }\textbf {\bibinfo {volume}
  {119}},\ \bibinfo {pages} {240401} (\bibinfo {year} {2017})}\BibitemShut
  {NoStop}%
\bibitem [{\citenamefont {Marletto}\ and\ \citenamefont
  {Vedral}(2017)}]{MarVed17}%
  \BibitemOpen
  \bibfield  {author} {\bibinfo {author} {\bibfnamefont {C.}~\bibnamefont
  {Marletto}}\ and\ \bibinfo {author} {\bibfnamefont {V.}~\bibnamefont
  {Vedral}},\ }\bibfield  {title} {\bibinfo {title} {Gravitationally {Induced}
  {Entanglement} between {Two} {Massive} {Particles} is {Sufficient} {Evidence}
  of {Quantum} {Effects} in {Gravity}},\ }\href
  {https://doi.org/10.1103/PhysRevLett.119.240402} {\bibfield  {journal}
  {\bibinfo  {journal} {Physical Review Letters}\ }\textbf {\bibinfo {volume}
  {119}},\ \bibinfo {pages} {240402} (\bibinfo {year} {2017})}\BibitemShut
  {NoStop}%
\bibitem [{\citenamefont {Aleksandrov}(1981)}]{Aleks81}%
  \BibitemOpen
  \bibfield  {author} {\bibinfo {author} {\bibfnamefont {I.~V.}\ \bibnamefont
  {Aleksandrov}},\ }\bibfield  {title} {\bibinfo {title} {The statistical
  dynamics of a system consisting of a classical and a quantum subsystem},\
  }\href {https://doi.org/doi:10.1515/zna-1981-0819} {\bibfield  {journal}
  {\bibinfo  {journal} {Zeitschrift f{\"u}r Naturforschung A}\ }\textbf
  {\bibinfo {volume} {36}},\ \bibinfo {pages} {902} (\bibinfo {year}
  {1981})}\BibitemShut {NoStop}%
\bibitem [{\citenamefont {Boucher}\ and\ \citenamefont
  {Traschen}(1988)}]{BouTras88}%
  \BibitemOpen
  \bibfield  {author} {\bibinfo {author} {\bibfnamefont {W.}~\bibnamefont
  {Boucher}}\ and\ \bibinfo {author} {\bibfnamefont {J.}~\bibnamefont
  {Traschen}},\ }\bibfield  {title} {\bibinfo {title} {Semiclassical physics
  and quantum fluctuations},\ }\href {https://doi.org/10.1103/PhysRevD.37.3522}
  {\bibfield  {journal} {\bibinfo  {journal} {Phys. Rev. D}\ }\textbf {\bibinfo
  {volume} {37}},\ \bibinfo {pages} {3522} (\bibinfo {year}
  {1988})}\BibitemShut {NoStop}%
\bibitem [{\citenamefont {Prezhdo}\ and\ \citenamefont
  {Kisil}(1997)}]{PrezKis97}%
  \BibitemOpen
  \bibfield  {author} {\bibinfo {author} {\bibfnamefont {O.~V.}\ \bibnamefont
  {Prezhdo}}\ and\ \bibinfo {author} {\bibfnamefont {V.~V.}\ \bibnamefont
  {Kisil}},\ }\bibfield  {title} {\bibinfo {title} {Mixing quantum and
  classical mechanics},\ }\href {https://doi.org/10.1103/PhysRevA.56.162}
  {\bibfield  {journal} {\bibinfo  {journal} {Phys. Rev. A}\ }\textbf {\bibinfo
  {volume} {56}},\ \bibinfo {pages} {162} (\bibinfo {year} {1997})}\BibitemShut
  {NoStop}%
\bibitem [{\citenamefont {Peres}\ and\ \citenamefont {Terno}(2001)}]{PerTer01}%
  \BibitemOpen
  \bibfield  {author} {\bibinfo {author} {\bibfnamefont {A.}~\bibnamefont
  {Peres}}\ and\ \bibinfo {author} {\bibfnamefont {D.~R.}\ \bibnamefont
  {Terno}},\ }\bibfield  {title} {\bibinfo {title} {Hybrid classical-quantum
  dynamics},\ }\href {https://doi.org/10.1103/PhysRevA.63.022101} {\bibfield
  {journal} {\bibinfo  {journal} {Phys. Rev. A}\ }\textbf {\bibinfo {volume}
  {63}},\ \bibinfo {pages} {022101} (\bibinfo {year} {2001})}\BibitemShut
  {NoStop}%
\bibitem [{\citenamefont {Anastopoulos}\ and\ \citenamefont
  {Hu}(2014)}]{AnHu14}%
  \BibitemOpen
  \bibfield  {author} {\bibinfo {author} {\bibfnamefont {C.}~\bibnamefont
  {Anastopoulos}}\ and\ \bibinfo {author} {\bibfnamefont {B.-L.}\ \bibnamefont
  {Hu}},\ }\bibfield  {title} {\bibinfo {title} {Problems with the
  {Newton}--{Schr{\"o}dinger} equations},\ }\href
  {https://doi.org/10.1088/1367-2630/16/8/085007} {\bibfield  {journal}
  {\bibinfo  {journal} {New Journal of Physics}\ }\textbf {\bibinfo {volume}
  {16}},\ \bibinfo {pages} {085007} (\bibinfo {year} {2014})}\BibitemShut
  {NoStop}%
\bibitem [{\citenamefont {Pound}\ and\ \citenamefont {Rebka}(1960)}]{PouReb60}%
  \BibitemOpen
  \bibfield  {author} {\bibinfo {author} {\bibfnamefont {R.~V.}\ \bibnamefont
  {Pound}}\ and\ \bibinfo {author} {\bibfnamefont {G.~A.}\ \bibnamefont
  {Rebka}},\ }\bibfield  {title} {\bibinfo {title} {Apparent {Weight} of
  {Photons}},\ }\href {https://doi.org/10.1103/PhysRevLett.4.337} {\bibfield
  {journal} {\bibinfo  {journal} {Physical Review Letters}\ }\textbf {\bibinfo
  {volume} {4}},\ \bibinfo {pages} {337} (\bibinfo {year} {1960})}\BibitemShut
  {NoStop}%
\bibitem [{\citenamefont {Colella}\ \emph {et~al.}(1975)\citenamefont
  {Colella}, \citenamefont {Overhauser},\ and\ \citenamefont {Werner}}]{cow75}%
  \BibitemOpen
  \bibfield  {author} {\bibinfo {author} {\bibfnamefont {R.}~\bibnamefont
  {Colella}}, \bibinfo {author} {\bibfnamefont {A.~W.}\ \bibnamefont
  {Overhauser}},\ and\ \bibinfo {author} {\bibfnamefont {S.~A.}\ \bibnamefont
  {Werner}},\ }\bibfield  {title} {\bibinfo {title} {Observation of
  {Gravitationally} {Induced} {Quantum} {Interference}},\ }\href
  {https://doi.org/10.1103/PhysRevLett.34.1472} {\bibfield  {journal} {\bibinfo
   {journal} {Physical Review Letters}\ }\textbf {\bibinfo {volume} {34}},\
  \bibinfo {pages} {1472} (\bibinfo {year} {1975})}\BibitemShut {NoStop}%
\bibitem [{\citenamefont {Peters}\ \emph {et~al.}(2001)\citenamefont {Peters},
  \citenamefont {Chung},\ and\ \citenamefont {Chu}}]{PetChu01}%
  \BibitemOpen
  \bibfield  {author} {\bibinfo {author} {\bibfnamefont {A.}~\bibnamefont
  {Peters}}, \bibinfo {author} {\bibfnamefont {K.~Y.}\ \bibnamefont {Chung}},\
  and\ \bibinfo {author} {\bibfnamefont {S.}~\bibnamefont {Chu}},\ }\bibfield
  {title} {\bibinfo {title} {High-precision gravity measurements using atom
  interferometry},\ }\href {https://doi.org/10.1088/0026-1394/38/1/4}
  {\bibfield  {journal} {\bibinfo  {journal} {Metrologia}\ }\textbf {\bibinfo
  {volume} {38}},\ \bibinfo {pages} {25} (\bibinfo {year} {2001})}\BibitemShut
  {NoStop}%
\bibitem [{\citenamefont {Nesvizhevsky}\ \emph {et~al.}(2002)\citenamefont
  {Nesvizhevsky}, \citenamefont {B{\"o}rner}, \citenamefont {Petukhov},
  \citenamefont {Abele}, \citenamefont {Bae{\ss}ler}, \citenamefont {Rue{\ss}},
  \citenamefont {St{\"o}ferle}, \citenamefont {Westphal}, \citenamefont
  {Gagarski}, \citenamefont {Petrov},\ and\ \citenamefont {Strelkov}}]{Nesv02}%
  \BibitemOpen
  \bibfield  {author} {\bibinfo {author} {\bibfnamefont {V.~V.}\ \bibnamefont
  {Nesvizhevsky}}, \bibinfo {author} {\bibfnamefont {H.~G.}\ \bibnamefont
  {B{\"o}rner}}, \bibinfo {author} {\bibfnamefont {A.~K.}\ \bibnamefont
  {Petukhov}}, \bibinfo {author} {\bibfnamefont {H.}~\bibnamefont {Abele}},
  \bibinfo {author} {\bibfnamefont {S.}~\bibnamefont {Bae{\ss}ler}}, \bibinfo
  {author} {\bibfnamefont {F.~J.}\ \bibnamefont {Rue{\ss}}}, \bibinfo {author}
  {\bibfnamefont {T.}~\bibnamefont {St{\"o}ferle}}, \bibinfo {author}
  {\bibfnamefont {A.}~\bibnamefont {Westphal}}, \bibinfo {author}
  {\bibfnamefont {A.~M.}\ \bibnamefont {Gagarski}}, \bibinfo {author}
  {\bibfnamefont {G.~A.}\ \bibnamefont {Petrov}},\ and\ \bibinfo {author}
  {\bibfnamefont {A.~V.}\ \bibnamefont {Strelkov}},\ }\bibfield  {title}
  {\bibinfo {title} {Quantum states of neutrons in the {Earth}'s gravitational
  field},\ }\href {https://doi.org/10.1038/415297a} {\bibfield  {journal}
  {\bibinfo  {journal} {Nature}\ }\textbf {\bibinfo {volume} {415}},\ \bibinfo
  {pages} {297} (\bibinfo {year} {2002})}\BibitemShut {NoStop}%
\bibitem [{\citenamefont {Jenke}\ \emph {et~al.}(2011)\citenamefont {Jenke},
  \citenamefont {Geltenbort}, \citenamefont {Lemmel},\ and\ \citenamefont
  {Abele}}]{Jenk11}%
  \BibitemOpen
  \bibfield  {author} {\bibinfo {author} {\bibfnamefont {T.}~\bibnamefont
  {Jenke}}, \bibinfo {author} {\bibfnamefont {P.}~\bibnamefont {Geltenbort}},
  \bibinfo {author} {\bibfnamefont {H.}~\bibnamefont {Lemmel}},\ and\ \bibinfo
  {author} {\bibfnamefont {H.}~\bibnamefont {Abele}},\ }\bibfield  {title}
  {\bibinfo {title} {Realization of a gravity-resonance-spectroscopy
  technique},\ }\href {https://doi.org/10.1038/nphys1970} {\bibfield  {journal}
  {\bibinfo  {journal} {Nature Physics}\ }\textbf {\bibinfo {volume} {7}},\
  \bibinfo {pages} {468} (\bibinfo {year} {2011})}\BibitemShut {NoStop}%
\bibitem [{\citenamefont {Rosi}\ \emph {et~al.}(2014)\citenamefont {Rosi},
  \citenamefont {Sorrentino}, \citenamefont {Cacciapuoti}, \citenamefont
  {Prevedelli},\ and\ \citenamefont {Tino}}]{RosTino14}%
  \BibitemOpen
  \bibfield  {author} {\bibinfo {author} {\bibfnamefont {G.}~\bibnamefont
  {Rosi}}, \bibinfo {author} {\bibfnamefont {F.}~\bibnamefont {Sorrentino}},
  \bibinfo {author} {\bibfnamefont {L.}~\bibnamefont {Cacciapuoti}}, \bibinfo
  {author} {\bibfnamefont {M.}~\bibnamefont {Prevedelli}},\ and\ \bibinfo
  {author} {\bibfnamefont {G.~M.}\ \bibnamefont {Tino}},\ }\bibfield  {title}
  {\bibinfo {title} {Precision measurement of the {Newtonian} gravitational
  constant using cold atoms},\ }\href {https://doi.org/10.1038/nature13433}
  {\bibfield  {journal} {\bibinfo  {journal} {Nature}\ }\textbf {\bibinfo
  {volume} {510}},\ \bibinfo {pages} {518} (\bibinfo {year}
  {2014})}\BibitemShut {NoStop}%
\bibitem [{\citenamefont {Asenbaum}\ \emph {et~al.}(2017)\citenamefont
  {Asenbaum}, \citenamefont {Overstreet}, \citenamefont {Kovachy},
  \citenamefont {Brown}, \citenamefont {Hogan},\ and\ \citenamefont
  {Kasevich}}]{AseKas17}%
  \BibitemOpen
  \bibfield  {author} {\bibinfo {author} {\bibfnamefont {P.}~\bibnamefont
  {Asenbaum}}, \bibinfo {author} {\bibfnamefont {C.}~\bibnamefont
  {Overstreet}}, \bibinfo {author} {\bibfnamefont {T.}~\bibnamefont {Kovachy}},
  \bibinfo {author} {\bibfnamefont {D.~D.}\ \bibnamefont {Brown}}, \bibinfo
  {author} {\bibfnamefont {J.~M.}\ \bibnamefont {Hogan}},\ and\ \bibinfo
  {author} {\bibfnamefont {M.~A.}\ \bibnamefont {Kasevich}},\ }\bibfield
  {title} {\bibinfo {title} {Phase {Shift} in an {Atom} {Interferometer} due to
  {Spacetime} {Curvature} across its {Wave} {Function}},\ }\href
  {https://doi.org/10.1103/PhysRevLett.118.183602} {\bibfield  {journal}
  {\bibinfo  {journal} {Physical Review Letters}\ }\textbf {\bibinfo {volume}
  {118}},\ \bibinfo {pages} {183602} (\bibinfo {year} {2017})}\BibitemShut
  {NoStop}%
\bibitem [{\citenamefont {Dirac}(1950)}]{Dirac_1950}%
  \BibitemOpen
  \bibfield  {author} {\bibinfo {author} {\bibfnamefont {P.~A.~M.}\
  \bibnamefont {Dirac}},\ }\bibfield  {title} {\bibinfo {title} {Generalized
  hamiltonian dynamics},\ }\href {https://doi.org/10.4153/CJM-1950-012-1}
  {\bibfield  {journal} {\bibinfo  {journal} {Canadian Journal of Mathematics}\
  }\textbf {\bibinfo {volume} {2}},\ \bibinfo {pages} {129} (\bibinfo {year}
  {1950})}\BibitemShut {NoStop}%
\bibitem [{\citenamefont {Dirac}(1958)}]{Dirac58}%
  \BibitemOpen
  \bibfield  {author} {\bibinfo {author} {\bibfnamefont {P.~A.~M.}\
  \bibnamefont {Dirac}},\ }\bibfield  {title} {\bibinfo {title} {Generalized
  {Hamiltonian} {Dynamics}},\ }\href {https://doi.org/10.1098/rspa.1958.0141}
  {\bibfield  {journal} {\bibinfo  {journal} {Proceedings of the Royal Society
  A}\ }\textbf {\bibinfo {volume} {246}},\ \bibinfo {pages} {326} (\bibinfo
  {year} {1958})}\BibitemShut {NoStop}%
\bibitem [{\citenamefont {Di{\'o}si}(2011)}]{Di_si_2011}%
  \BibitemOpen
  \bibfield  {author} {\bibinfo {author} {\bibfnamefont {L.}~\bibnamefont
  {Di{\'o}si}},\ }\bibfield  {title} {\bibinfo {title} {The gravity-related
  decoherence master equation from hybrid dynamics},\ }\href
  {https://doi.org/10.1088/1742-6596/306/1/012006} {\bibfield  {journal}
  {\bibinfo  {journal} {Journal of Physics: Conference Series}\ }\textbf
  {\bibinfo {volume} {306}},\ \bibinfo {pages} {012006} (\bibinfo {year}
  {2011})}\BibitemShut {NoStop}%
\bibitem [{\citenamefont {Carroll}(2003)}]{carroll2003spacetime}%
  \BibitemOpen
  \bibfield  {author} {\bibinfo {author} {\bibfnamefont {S.}~\bibnamefont
  {Carroll}},\ }\href
  {http://www.amazon.com/Spacetime-Geometry-Introduction-General-Relativity/dp/0805387323}
  {\emph {\bibinfo {title} {Spacetime and Geometry: An Introduction to General
  Relativity}}}\ (\bibinfo  {publisher} {Benjamin Cummings},\ \bibinfo {year}
  {2003})\BibitemShut {NoStop}%
\bibitem [{\citenamefont {Dirac}(1959)}]{Dir59}%
  \BibitemOpen
  \bibfield  {author} {\bibinfo {author} {\bibfnamefont {P.~A.~M.}\
  \bibnamefont {Dirac}},\ }\bibfield  {title} {\bibinfo {title} {Fixation of
  {Coordinates} in the {Hamiltonian} {Theory} of {Gravitation}},\ }\href
  {https://doi.org/10.1103/PhysRev.114.924} {\bibfield  {journal} {\bibinfo
  {journal} {Physical Review}\ }\textbf {\bibinfo {volume} {114}},\ \bibinfo
  {pages} {924} (\bibinfo {year} {1959})}\BibitemShut {NoStop}%
\bibitem [{\citenamefont {Arnowitt}\ \emph {et~al.}(1959)\citenamefont
  {Arnowitt}, \citenamefont {Deser},\ and\ \citenamefont {Misner}}]{ADM59}%
  \BibitemOpen
  \bibfield  {author} {\bibinfo {author} {\bibfnamefont {R.}~\bibnamefont
  {Arnowitt}}, \bibinfo {author} {\bibfnamefont {S.}~\bibnamefont {Deser}},\
  and\ \bibinfo {author} {\bibfnamefont {C.~W.}\ \bibnamefont {Misner}},\
  }\bibfield  {title} {\bibinfo {title} {Dynamical {Structure} and {Definition}
  of {Energy} in {General} {Relativity}},\ }\href
  {https://doi.org/10.1103/PhysRev.116.1322} {\bibfield  {journal} {\bibinfo
  {journal} {Physical Review}\ }\textbf {\bibinfo {volume} {116}},\ \bibinfo
  {pages} {1322} (\bibinfo {year} {1959})}\BibitemShut {NoStop}%
\bibitem [{Note1()}]{Note1}%
  \BibitemOpen
  \bibinfo {note} {Other gauge choices, such as in Ref. \cite {LayOpp23}, give
  rise to other degrees of freedom; however in the weak-field adiabatic limit,
  agreement between the different approaches is manifest.}\BibitemShut {Stop}%
\bibitem [{\citenamefont {Layton}\ \emph {et~al.}(2023)\citenamefont {Layton},
  \citenamefont {Oppenheim}, \citenamefont {Russo},\ and\ \citenamefont
  {Weller-Davies}}]{LayOpp23}%
  \BibitemOpen
  \bibfield  {author} {\bibinfo {author} {\bibfnamefont {I.}~\bibnamefont
  {Layton}}, \bibinfo {author} {\bibfnamefont {J.}~\bibnamefont {Oppenheim}},
  \bibinfo {author} {\bibfnamefont {A.}~\bibnamefont {Russo}},\ and\ \bibinfo
  {author} {\bibfnamefont {Z.}~\bibnamefont {Weller-Davies}},\ }\bibfield
  {title} {\bibinfo {title} {The weak field limit of quantum matter
  back-reacting on classical spacetime},\ }\href
  {https://doi.org/10.1007/JHEP08(2023)163} {\bibfield  {journal} {\bibinfo
  {journal} {Journal of High Energy Physics}\ }\textbf {\bibinfo {volume}
  {2023}},\ \bibinfo {pages} {163} (\bibinfo {year} {2023})}\BibitemShut
  {NoStop}%
\bibitem [{Note2()}]{Note2}%
  \BibitemOpen
  \bibinfo {note} {This means that the constraints derived in Refs. \cite
  {OppZach22,LayOpp23}, based on positive-definiteness of the evolution, are
  inapplicable to $\epsilon $.}\BibitemShut {Stop}%
\bibitem [{\citenamefont {Anastopoulos}\ \emph {et~al.}(2021)\citenamefont
  {Anastopoulos}, \citenamefont {Lagouvardos},\ and\ \citenamefont
  {Savvidou}}]{Anastopoulos_2021}%
  \BibitemOpen
  \bibfield  {author} {\bibinfo {author} {\bibfnamefont {C.}~\bibnamefont
  {Anastopoulos}}, \bibinfo {author} {\bibfnamefont {M.}~\bibnamefont
  {Lagouvardos}},\ and\ \bibinfo {author} {\bibfnamefont {K.}~\bibnamefont
  {Savvidou}},\ }\bibfield  {title} {\bibinfo {title} {Gravitational effects in
  macroscopic quantum systems: a first-principles analysis},\ }\href
  {https://doi.org/10.1088/1361-6382/ac0bf9} {\bibfield  {journal} {\bibinfo
  {journal} {Classical and Quantum Gravity}\ }\textbf {\bibinfo {volume}
  {38}},\ \bibinfo {pages} {155012} (\bibinfo {year} {2021})}\BibitemShut
  {NoStop}%
\bibitem [{\citenamefont {Gardiner}\ and\ \citenamefont
  {Zoller}(2000)}]{gardiner00}%
  \BibitemOpen
  \bibfield  {author} {\bibinfo {author} {\bibfnamefont {C.~W.}\ \bibnamefont
  {Gardiner}}\ and\ \bibinfo {author} {\bibfnamefont {P.}~\bibnamefont
  {Zoller}},\ }\href@noop {} {\emph {\bibinfo {title} {Quantum Noise}}},\
  \bibinfo {edition} {2nd}\ ed.,\ edited by\ \bibinfo {editor} {\bibfnamefont
  {H.}~\bibnamefont {Haken}}\ (\bibinfo  {publisher} {Springer},\ \bibinfo
  {year} {2000})\BibitemShut {NoStop}%
\bibitem [{\citenamefont {Braginsky}\ and\ \citenamefont
  {Khalili}(1992)}]{BragKhal92}%
  \BibitemOpen
  \bibfield  {author} {\bibinfo {author} {\bibfnamefont {V.~B.}\ \bibnamefont
  {Braginsky}}\ and\ \bibinfo {author} {\bibfnamefont {F.~Y.}\ \bibnamefont
  {Khalili}},\ }\href@noop {} {\emph {\bibinfo {title} {Quantum Measurement}}}\
  (\bibinfo  {publisher} {Cambridge University Press},\ \bibinfo {year}
  {1992})\BibitemShut {NoStop}%
\bibitem [{\citenamefont {Purdy}\ \emph {et~al.}(2013)\citenamefont {Purdy},
  \citenamefont {Peterson},\ and\ \citenamefont {Regal}}]{PurReg13a}%
  \BibitemOpen
  \bibfield  {author} {\bibinfo {author} {\bibfnamefont {T.~P.}\ \bibnamefont
  {Purdy}}, \bibinfo {author} {\bibfnamefont {R.~W.}\ \bibnamefont
  {Peterson}},\ and\ \bibinfo {author} {\bibfnamefont {C.~A.}\ \bibnamefont
  {Regal}},\ }\bibfield  {title} {\bibinfo {title} {Observation of {Radiation}
  {Pressure} {Shot} {Noise} on a {Macroscopic} {Object}},\ }\href
  {https://doi.org/10.1126/science.1231282} {\bibfield  {journal} {\bibinfo
  {journal} {Science}\ }\textbf {\bibinfo {volume} {339}},\ \bibinfo {pages}
  {801} (\bibinfo {year} {2013})}\BibitemShut {NoStop}%
\bibitem [{\citenamefont {Teufel}\ \emph {et~al.}(2016)\citenamefont {Teufel},
  \citenamefont {Lecocq},\ and\ \citenamefont {Simmonds}}]{Teuf16}%
  \BibitemOpen
  \bibfield  {author} {\bibinfo {author} {\bibfnamefont {J.}~\bibnamefont
  {Teufel}}, \bibinfo {author} {\bibfnamefont {F.}~\bibnamefont {Lecocq}},\
  and\ \bibinfo {author} {\bibfnamefont {R.}~\bibnamefont {Simmonds}},\
  }\bibfield  {title} {\bibinfo {title} {Overwhelming {Thermomechanical}
  {Motion} with {Microwave} {Radiation} {Pressure} {Shot} {Noise}},\ }\href
  {https://doi.org/10.1103/PhysRevLett.116.013602} {\bibfield  {journal}
  {\bibinfo  {journal} {Physical Review Letters}\ }\textbf {\bibinfo {volume}
  {116}},\ \bibinfo {pages} {013602} (\bibinfo {year} {2016})}\BibitemShut
  {NoStop}%
\bibitem [{\citenamefont {Suh}\ \emph {et~al.}(2014)\citenamefont {Suh},
  \citenamefont {Weinstein}, \citenamefont {Lei}, \citenamefont {Wollman},
  \citenamefont {Steinke}, \citenamefont {Meystre}, \citenamefont {Clerk},\
  and\ \citenamefont {Schwab}}]{SuSchw14}%
  \BibitemOpen
  \bibfield  {author} {\bibinfo {author} {\bibfnamefont {J.}~\bibnamefont
  {Suh}}, \bibinfo {author} {\bibfnamefont {A.~J.}\ \bibnamefont {Weinstein}},
  \bibinfo {author} {\bibfnamefont {C.~U.}\ \bibnamefont {Lei}}, \bibinfo
  {author} {\bibfnamefont {E.~E.}\ \bibnamefont {Wollman}}, \bibinfo {author}
  {\bibfnamefont {S.~K.}\ \bibnamefont {Steinke}}, \bibinfo {author}
  {\bibfnamefont {P.}~\bibnamefont {Meystre}}, \bibinfo {author} {\bibfnamefont
  {A.~A.}\ \bibnamefont {Clerk}},\ and\ \bibinfo {author} {\bibfnamefont
  {K.~C.}\ \bibnamefont {Schwab}},\ }\bibfield  {title} {\bibinfo {title}
  {Mechanically detecting and avoiding the quantum fluctuations of a microwave
  field},\ }\href {https://doi.org/10.1126/science.1253258} {\bibfield
  {journal} {\bibinfo  {journal} {Science}\ }\textbf {\bibinfo {volume}
  {344}},\ \bibinfo {pages} {1262} (\bibinfo {year} {2014})}\BibitemShut
  {NoStop}%
\bibitem [{\citenamefont {Lecocq}\ \emph {et~al.}(2015)\citenamefont {Lecocq},
  \citenamefont {Clark}, \citenamefont {Simmonds}, \citenamefont {Aumentado},\
  and\ \citenamefont {Teufel}}]{LecTeu15}%
  \BibitemOpen
  \bibfield  {author} {\bibinfo {author} {\bibfnamefont {F.}~\bibnamefont
  {Lecocq}}, \bibinfo {author} {\bibfnamefont {J.}~\bibnamefont {Clark}},
  \bibinfo {author} {\bibfnamefont {R.}~\bibnamefont {Simmonds}}, \bibinfo
  {author} {\bibfnamefont {J.}~\bibnamefont {Aumentado}},\ and\ \bibinfo
  {author} {\bibfnamefont {J.}~\bibnamefont {Teufel}},\ }\bibfield  {title}
  {\bibinfo {title} {Quantum {Nondemolition} {Measurement} of a {Nonclassical}
  {State} of a {Massive} {Object}},\ }\href
  {https://doi.org/10.1103/PhysRevX.5.041037} {\bibfield  {journal} {\bibinfo
  {journal} {Physical Review X}\ }\textbf {\bibinfo {volume} {5}},\ \bibinfo
  {pages} {041037} (\bibinfo {year} {2015})}\BibitemShut {NoStop}%
\bibitem [{\citenamefont {Yap}\ \emph {et~al.}(2019)\citenamefont {Yap},
  \citenamefont {Cripe}, \citenamefont {Mansell}, \citenamefont {McRae},
  \citenamefont {Ward}, \citenamefont {Slagmolen}, \citenamefont {Heu},
  \citenamefont {Follman}, \citenamefont {Cole}, \citenamefont {Corbitt},\ and\
  \citenamefont {McClelland}}]{YapCorb19}%
  \BibitemOpen
  \bibfield  {author} {\bibinfo {author} {\bibfnamefont {M.~J.}\ \bibnamefont
  {Yap}}, \bibinfo {author} {\bibfnamefont {J.}~\bibnamefont {Cripe}}, \bibinfo
  {author} {\bibfnamefont {G.~L.}\ \bibnamefont {Mansell}}, \bibinfo {author}
  {\bibfnamefont {T.~G.}\ \bibnamefont {McRae}}, \bibinfo {author}
  {\bibfnamefont {R.~L.}\ \bibnamefont {Ward}}, \bibinfo {author}
  {\bibfnamefont {B.~J.~J.}\ \bibnamefont {Slagmolen}}, \bibinfo {author}
  {\bibfnamefont {P.}~\bibnamefont {Heu}}, \bibinfo {author} {\bibfnamefont
  {D.}~\bibnamefont {Follman}}, \bibinfo {author} {\bibfnamefont {G.~D.}\
  \bibnamefont {Cole}}, \bibinfo {author} {\bibfnamefont {T.}~\bibnamefont
  {Corbitt}},\ and\ \bibinfo {author} {\bibfnamefont {D.~E.}\ \bibnamefont
  {McClelland}},\ }\bibfield  {title} {\bibinfo {title} {Broadband reduction of
  quantum radiation pressure noise via squeezed light injection},\ }\href
  {https://doi.org/10.1038/s41566-019-0527-y} {\bibfield  {journal} {\bibinfo
  {journal} {Nature Photonics}\ }\textbf {\bibinfo {volume} {14}},\ \bibinfo
  {pages} {19} (\bibinfo {year} {2019})}\BibitemShut {NoStop}%
\bibitem [{\citenamefont {Cahill}\ and\ \citenamefont
  {Glauber}(1969)}]{CahGlau69o}%
  \BibitemOpen
  \bibfield  {author} {\bibinfo {author} {\bibfnamefont {K.}~\bibnamefont
  {Cahill}}\ and\ \bibinfo {author} {\bibfnamefont {R.}~\bibnamefont
  {Glauber}},\ }\bibfield  {title} {\bibinfo {title} {Ordered {Expansions} in
  {Boson} {Amplitude} {Operators}},\ }\href
  {https://doi.org/10.1103/PhysRev.177.1857} {\bibfield  {journal} {\bibinfo
  {journal} {Physical Review}\ }\textbf {\bibinfo {volume} {177}},\ \bibinfo
  {pages} {1857} (\bibinfo {year} {1969})}\BibitemShut {NoStop}%
\bibitem [{\citenamefont {Simon}(2000)}]{Simon00}%
  \BibitemOpen
  \bibfield  {author} {\bibinfo {author} {\bibfnamefont {R.}~\bibnamefont
  {Simon}},\ }\bibfield  {title} {\bibinfo {title} {Peres-{Horodecki}
  {Separability} {Criterion} for {Continuous} {Variable} {Systems}},\ }\href
  {https://doi.org/10.1103/PhysRevLett.84.2726} {\bibfield  {journal} {\bibinfo
   {journal} {Physical Review Letters}\ }\textbf {\bibinfo {volume} {84}},\
  \bibinfo {pages} {2726} (\bibinfo {year} {2000})}\BibitemShut {NoStop}%
\bibitem [{\citenamefont {Duan}\ \emph {et~al.}(2000)\citenamefont {Duan},
  \citenamefont {Giedke}, \citenamefont {Cirac},\ and\ \citenamefont
  {Zoller}}]{Duan00}%
  \BibitemOpen
  \bibfield  {author} {\bibinfo {author} {\bibfnamefont {L.-M.}\ \bibnamefont
  {Duan}}, \bibinfo {author} {\bibfnamefont {G.}~\bibnamefont {Giedke}},
  \bibinfo {author} {\bibfnamefont {J.~I.}\ \bibnamefont {Cirac}},\ and\
  \bibinfo {author} {\bibfnamefont {P.}~\bibnamefont {Zoller}},\ }\bibfield
  {title} {\bibinfo {title} {Inseparability criterion for continuous variable
  systems},\ }\href {https://doi.org/10.1103/PhysRevLett.84.2722} {\bibfield
  {journal} {\bibinfo  {journal} {Physical Review Letters}\ }\textbf {\bibinfo
  {volume} {84}},\ \bibinfo {pages} {2722} (\bibinfo {year}
  {2000})}\BibitemShut {NoStop}%
\bibitem [{\citenamefont {Lami}\ \emph {et~al.}(2024)\citenamefont {Lami},
  \citenamefont {Pedernales},\ and\ \citenamefont {Plenio}}]{Lami24}%
  \BibitemOpen
  \bibfield  {author} {\bibinfo {author} {\bibfnamefont {L.}~\bibnamefont
  {Lami}}, \bibinfo {author} {\bibfnamefont {J.~S.}\ \bibnamefont
  {Pedernales}},\ and\ \bibinfo {author} {\bibfnamefont {M.~B.}\ \bibnamefont
  {Plenio}},\ }\bibfield  {title} {\bibinfo {title} {Testing the {Quantumness}
  of {Gravity} without {Entanglement}},\ }\href
  {https://doi.org/10.1103/PhysRevX.14.021022} {\bibfield  {journal} {\bibinfo
  {journal} {Physical Review X}\ }\textbf {\bibinfo {volume} {14}},\ \bibinfo
  {pages} {021022} (\bibinfo {year} {2024})}\BibitemShut {NoStop}%
\end{thebibliography}%

\newpage
\phantom{I hate this bibliogrphy link}

\foreach \x in {1,...,12}
{
\clearpage
\includepdf[pages={\x},angle=0]{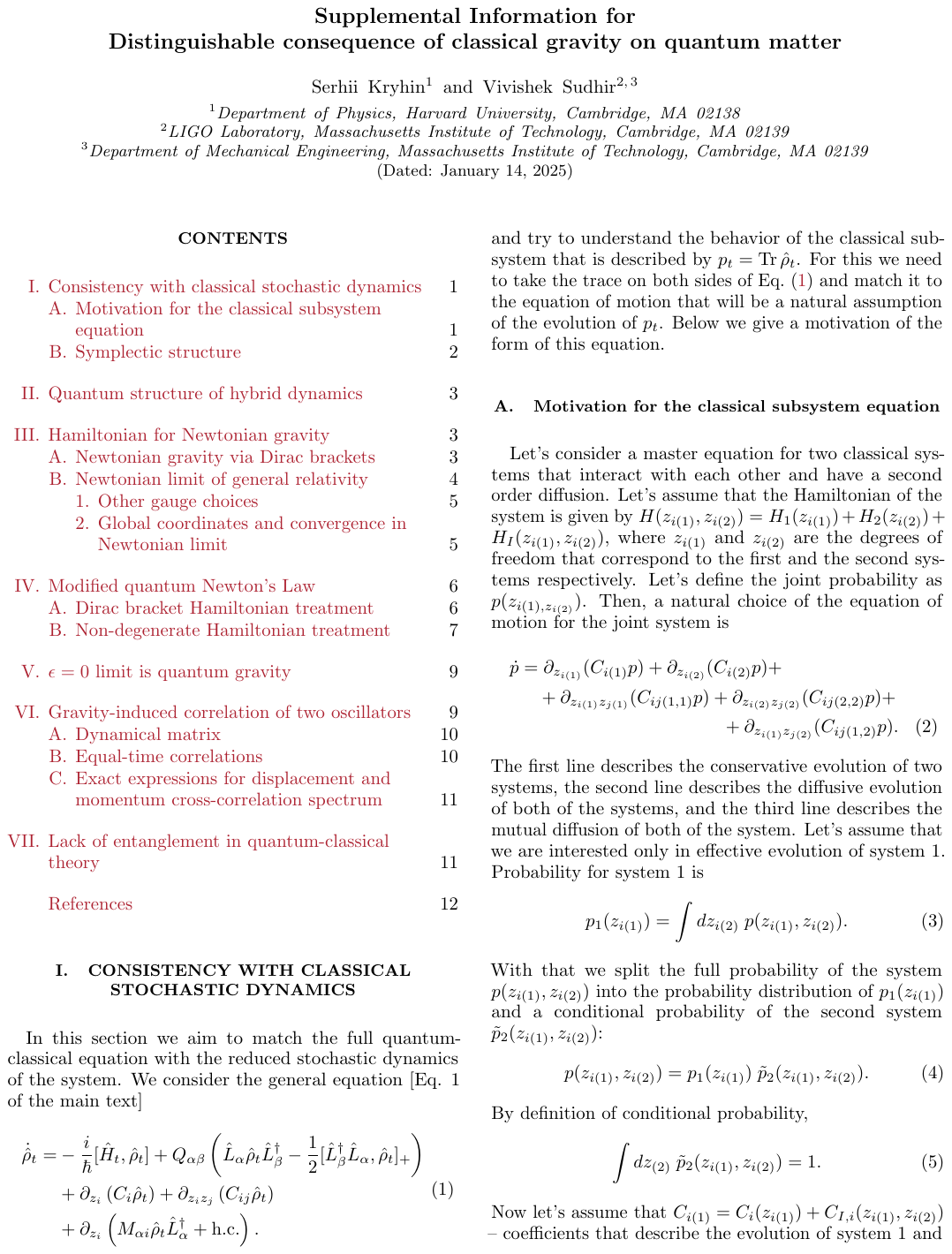} 
}

\end{document}